\documentclass{sig-alternate-05-2015}
\usepackage{times}
\usepackage[hide]{notes-alt}
\usepackage{amssymb}
\usepackage{amsmath}
\usepackage{graphicx}
\usepackage{subcaption}
\usepackage[ruled,vlined,linesnumberedhidden]{algorithm2e}
\usepackage{booktabs}
\usepackage{url}
\usepackage{listings}
\usepackage{color}
\usepackage[usenames,dvipsnames,svgnames,table]{xcolor}
\begin{document}
%
\CopyrightYear{2016}
\setcopyright{acmcopyright}
\conferenceinfo{CIKM'16 ,}{October 24-28, 2016, Indianapolis, IN, USA}
\isbn{978-1-4503-4073-1/16/10}\acmPrice{\$15.00}
\doi{http://dx.doi.org/10.1145/2983323.2983798}

\newcommand{\lm}{\textsc{Lm}}
\newcommand{\ideal}{\textsc{Ideal}}
\newcommand{\lmFeedback}{\textsc{Lm-Feedback}}
\newcommand{\interleavelmlm}{\textsc{I(Lm,Lm-Feedback)}}
\newcommand{\divEnt}{\textsc{$\text{Div}_{Ent}$}}
\newcommand{\divEntFeedback}{\textsc{$\text{Div}_{Ent}$-Feedback}}
\newcommand{\interleavedivEntEntFeedback}{\textsc{I($\text{Div}_{Ent}$, $\text{Div}_{Ent}$-Feedback)}}  
\newcommand{\interleaveLmDivEntFeedback}{\textsc{I(Lm, $\text{Div}_{Ent}$-Feedback)}}
\newcommand{\divKp}{\textsc{$\text{Div}_{Kp}$}}
\newcommand{\divKpFeedback}{\textsc{$\text{Div}_{Kp}$-Feedback}}
\newcommand{\interleavedivKpKpFeedback}{\textsc{I($\text{Div}_{Kp}$, $\text{Div}_{Kp}$-Feedback)}}
\newcommand{\interleaveLmDivKpFeedback}{\textsc{I(Lm, $\text{Div}_{Kp}$-Feedback)}}

\title{Discovering Entities with Just a Little Help from You}
%
%
%
%
%

\numberofauthors{3} 
%
%
%

\author{
\alignauthor Jaspreet Singh\\
        \affaddr{L3S Research Center}\\
        \affaddr{Leibniz Universit\"at Hannover}\\
        \email{singh@L3S.de}
\alignauthor Johannes Hoffart\\
       \affaddr{Max Planck Institute for Informatics}\\
       \email{jhoffart@mpi-inf.mpg.de}
\alignauthor Avishek Anand\\
        \affaddr{L3S Research Center}\\
        \affaddr{Leibniz Universit\"at Hannover}\\
        \email{anand@L3S.de}
}



\maketitle
\begin{abstract}
\vspace{3pt}
Linking entities like people, organizations, books, music groups and their songs in text to knowledge bases (KBs) is a
fundamental task for many downstream search and mining applications. Achieving high disambiguation accuracy crucially depends on a rich and holistic representation of the entities in the KB. For popular entities, such a representation can be easily mined from Wikipedia, and many current entity disambiguation and linking methods make use of this fact. However, Wikipedia does not contain long-tail entities that only few people are interested in, and also at times lags behind until newly emerging entities are added. For such entities, mining a suitable representation in a fully automated fashion is very difficult, resulting in poor linking accuracy.


What can automatically be mined, though, is a high-quality representation given the context of a new entity occurring in any text. Due to the lack of knowledge about the entity, no method can retrieve these occurrences automatically with high precision, resulting in a chicken-egg problem. To address this, our approach automatically generates candidate occurrences of entities, prompting the user for feedback to decide if the occurrence refers to the actual entity in question. This feedback gradually improves the knowledge and allows our methods to provide better candidate suggestions to keep the user engaged. We propose novel human-in-the-loop retrieval methods for generating candidates based on gradient interleaving of diversification and textual relevance approaches. 

We conducted extensive experiments on the FACC dataset, showing that our approaches convincingly outperform carefully selected baselines in both intrinsic and extrinsic measures while keeping users engaged.
\vspace{3pt}

\end{abstract}





\section{Introduction}
\vspace{5pt}
\note{
\begin{enumerate}
	\item introduce NED and the task of disambiguation. explain briefly what a KB is and how its used in NED. what is an ambiguous entity?
	\item why do we need context (keyphrases) for entity disambiguation. right now context gathered automatically. 
	\item this is good for head entities - lots of newspaper articles also exist, people also put more into wikipedia for these entities
	\item what if the entitiy is not in the KB? long tail entities tend to be absent or under represented. why cant we use automatic methods for gathering context for long tail entities? - using external sources of text - lots of noise - hard for machines to pick the right context especially for ambiguous entities
	\item can we intvolve the user to add more long tail entities?
	\item Explain the scenario - user queries with mention and some keywords, show one document at a time
	\item why one document at a time?
	\item We focus on ambiguous long tail entities
	\item interface diagram
	\item Why is ranking documents in such a scenario difficult?
\end{enumerate}
}


\subsection{Motivation} 
Connecting texts to knowledge bases (KBs) by linking names to the KB's
canonical entities such as people, organizations, or movies and their characters, is
a fundamental first step for a broad range of applications. Beyond the tasks
of language understanding, question answering, and information extraction, one
key application that has recently emerged is the use of entities in
information retrieval. 

However, all of these applications crucially depend on the fact that all entities
of interest are present in the KB. This is problematic when dealing with long-tail entities, entities not prominent
enough to be included in general domain KBs yet important for domain-specific
search tasks, e.\,g. patent search or historical
search~\cite{singh:chiir2016}. 
The problem is equally acute
with emerging entities, i.\,e.\ entities that are completely new or are just gaining
popularity, as it often takes considerable time for entities to be added to
a KB~\cite{Keegan:2013gq,DBLP:conf/websci/FetahuAA15}.




{I}magine for example a Star Wars fan, who --- after seeing the latest movie ---
wants to know what others think of her favorite character
\textbf{Rey} by looking at social media. Searching for just the string
``Rey'' will turn up a lot of uninteresting results about other Reys,
e.\,g.\ the \texttt{Copa del Rey} or other people sharing the same name. However, the new movie character Rey might not (yet) be in the KB.


\subsection{Problem} 

Quickly identifying descriptions for emerging or long-tail
entities suitable for users to understand the entity and for
linking further texts to the new entity is thus the key problem. Such
descriptions are one of the fundamental building blocks for entity linking
methods~\cite{Shen:2015wi}, where they are used for computing the textual similarity
of an (ambiguous) name in a text and an entity in a KB.


Automated approaches which harvest entity representations from 
text collections typically depend on encyclopedic sources (like 
\\ Wikipedia) and thus 
suffer due to \emph{sparsity} for emerging and long-tail entities.
On the other extreme, using unsupervised methods for Web collections, with potentially 
higher recall, are not accurate enough~\cite{Hoffart:2014hp,TACKBP:2015}.

\subsection{Solution} The most promising way to achieve
human-like quality when adding entities is with
with the help of the user herself. However, even such manual curation
must be well supported by the system to avoid putting 
undue burden on the user. A straightforward way to obtain the description would be to ask the user for
phrases that are salient and descriptive for the entity to be added. However, this
would soon become boring for the user and result in poor keyphrases. Additionally, checking if the entity already exists in the KB
is cumbersome. When users lose attention and care, there is a high risk
of adding duplicate entities.

\begin{figure*}[t]
\centering
\parbox[c][3cm]{0.85\textwidth}{
$\dots$ \emph{Game maker Hasbro} will include female \emph{Star Wars: The Force Awakens} character \textbf{Rey} in their \emph{Star Wars} themed \emph{Monopoly} game $\dots$

\vspace*{2mm}

$\dots$ the new face of \emph{Star Wars} alongside his onscreen co-star \textbf{Rey}, played by \emph{Daisy Ridley} $\dots$

\vspace*{2mm}

$\dots$ promising \emph{newbies} include \textbf{Rey}, a \emph{tough loner} who lost her family and scavenges on the \emph{desert planet Jakku} $\dots$
}
\caption{Entities-in-Context (EICs) for the Star Wars character Rey (keyphrases are \emph{emphasized})}
\label{fig:eics}
\end{figure*}

\begin{figure*}[!ht]
  \centering
  \begin{subfigure}[b]{0.37\textwidth}
     \includegraphics[width=0.8\textwidth]{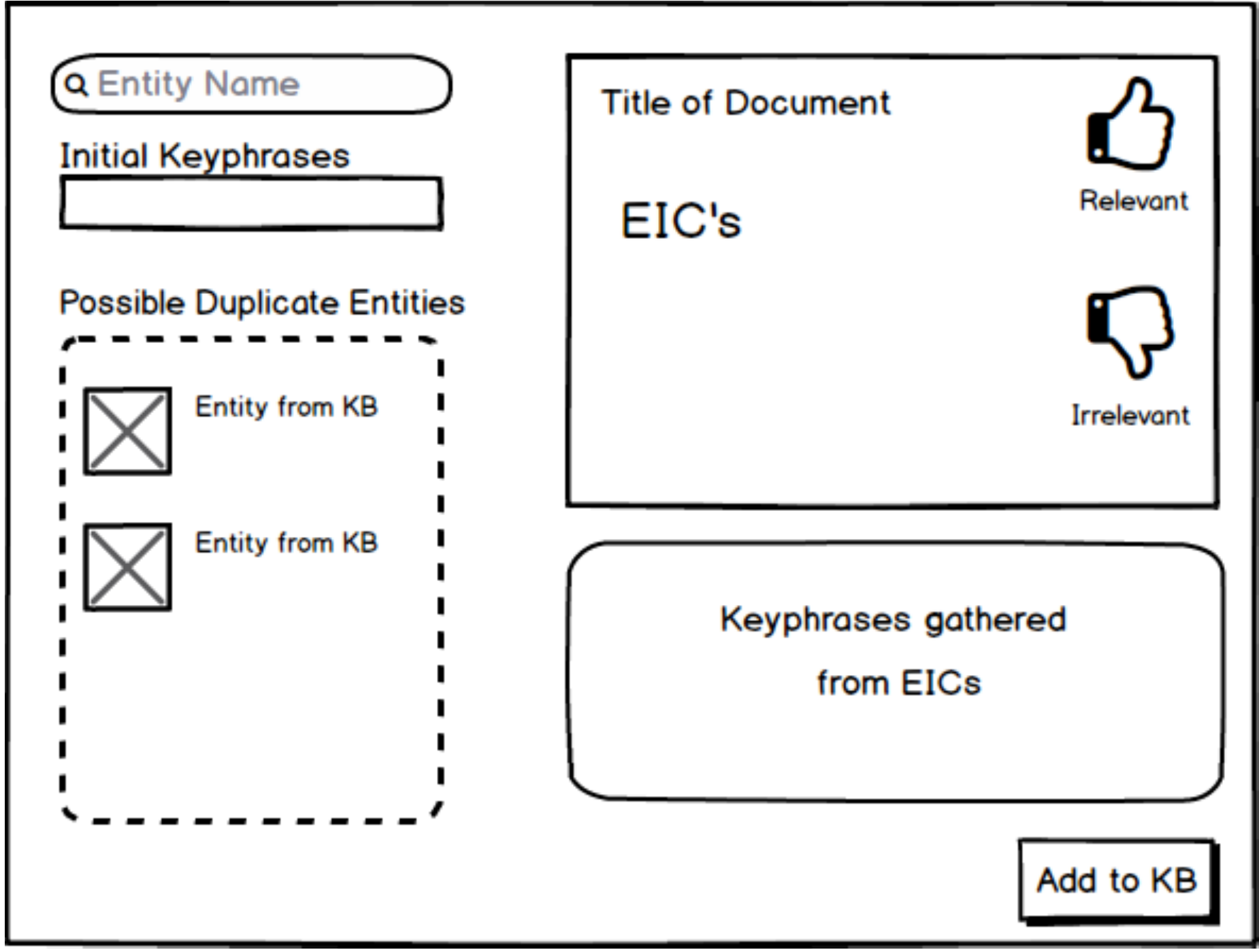}
    \caption{User Interface Mockup}
    \label{fig:ui}
  \end{subfigure}
  \begin{subfigure}[b]{0.59\textwidth}
     \includegraphics[width=0.96\textwidth]{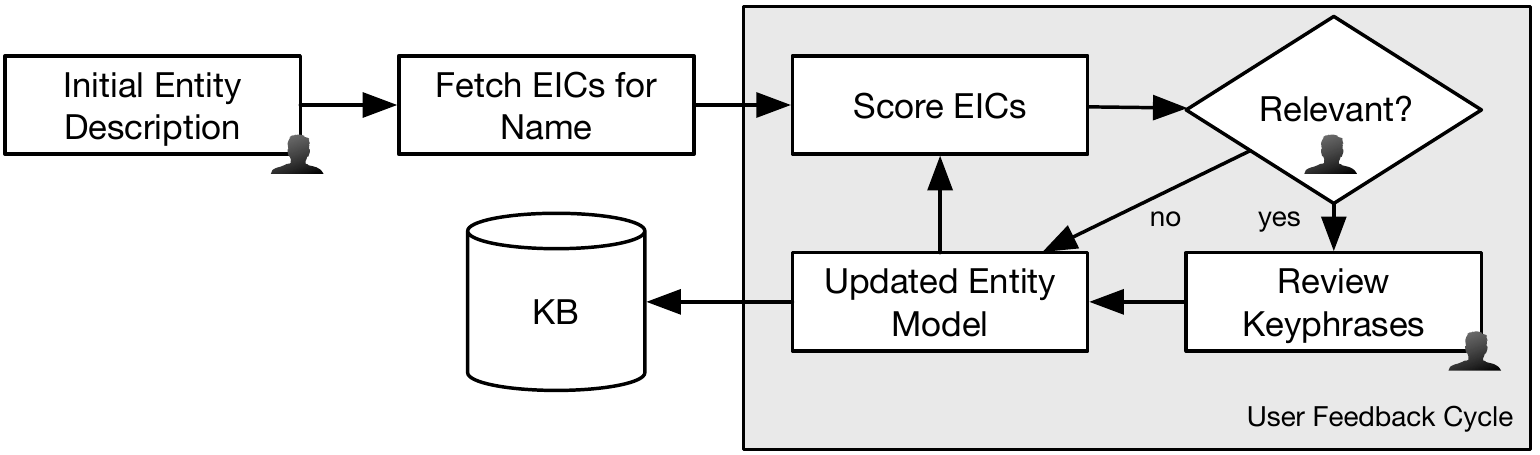}
    \caption{User flow through the system}
    \label{fig:userflow}
  \end{subfigure}
 
  \caption{Harvesting Keyphrases with the Help of the User}
  \label{fig:arch}
  \vspace{-7pt}
\end{figure*}

Our idea to keep the user engaged and motivated is to present her with entities-in-context (EICs, see Figure~1), i.\,e.\ snippets of
text containing the entity name and some context, which she merely has to accept or 
reject. We model the \emph{entity addition task} as a retrieval task where the user is
shown \emph{one document at a time}, documents that are likely containers of EICs, which she 
evaluates for relevance with respect to her intended query (input as keywords). This is a low-overhead activity, and
a lot of people do this in everyday applications, for example, 
to tag faces in photographs
(Apple's iPhoto comes to mind) or to find matching partners (Tinder). From
accepted EICs we can automatically distill keyphrases to create an on-the-fly
description for the new entity. A typical system created for this task is shown in Figure~\ref{fig:ui}.

The goal of the entity addition retrieval task is to provide the user with documents, one at a time, to help her arrive at a rich description that allows for high linking accuracy. Note that in this paper we focus on the addition of \textit{long-tail ambiguous} entities for which human-in-the-loop is essential. To this end, our paper makes the following novel contributions:


\begin{itemize}
	\item We devise methods that optimize for keyphrase coverage which serves as proxy for the hard-to-estimate expected linking accuracy for the new entity.
	\item Since in our scenario the user can abandon the task at any time, it is important to keep the user engaged to cover many keyphrases. We devise a metric known as the engagement index based on novelty specific to our scenario to measure user engagement. 
	\item Our methods incorporate user feedback obtained during the addition process to identify irrelevant aspects of the ambiguous entity and increase the fraction of relevant EICs shown to the user.
	\item We propose different diversification approaches to maximize the coverage of relevant keyphrases, avoiding narrowing down on a certain set of keyphrases too quickly. 
	\item Finally, we propose novel \emph{result list interleaving} methods that combines relevance feedback and diversification to maximize both keyphrase coverage and user engagement.
\end{itemize}

We conducted extensive experiments with real users and user simulations, showing that our approaches convincingly outperform carefully selected baselines in both intrinsic and extrinsic measures while keeping the users engaged.




\section{Adding Entities}
\label{sub:problem}



The key requirement for adding new entities is that the representation should
be suitable for disambiguating the entity in new texts. There is a large and
growing body of work on entity disambiguation
\cite{Shen:2015wi}, and many methods using different features have been
created over the past years.  Entity disambiguation methods commonly first identify all \emph{mentions} of
entities in a piece of text, i.\,e.\ all names of people, organizations,
movies, locations, etc. All candidate entities are retrieved from the
knowledge base based on the overlap of their names with the mention. Crucial features to decide the correct entity among all candidates are:

\begin{itemize}
	\item  the importance of an entity with respect to the KB (and sometimes the mention)
	\item  the coherence between entities in a single text
	\item  the \emph{textual description} of an entity
\end{itemize}

In principle, almost all features
can be mined from an entity-in-context. In this paper we focus on keyphrases, i.e., the
textual description of entities, as the central feature, which will be harvested
with the help of the user. The textual description is one of the core features for several entity disambiguation methods~\cite{Bunescu:2006,Ratinov:2011,Hoffart:2011a}.




First of all, the user provides a minimal description of the entity $e$ to be
added, consisting of the (ambiguous) name and optionally some initial keyphrases. Using
this, our methods retrieve documents based on the estimated relevance from a
document collection $\mathcal{D}$. These documents are presented to the user
in the form of EICs, as in this representation it is easier to judge the entity.
The user interface of an entity-addition system and the full user process is modelled as shown in Figure~\ref{fig:arch}, where
components requiring user feedback are marked with a small shape of a head.
The goal of our methods is to produce a ranking that:
\begin{itemize}
	\item covers keyphrases such that disambiguation accuracy for $e$
is high,
	\item engages the user.

\end{itemize}
Note, the user has full control
over the retrieval depth which is different from other ranking tasks where the
objective is to retrieve the top-$k$ documents with the assumption she will
evaluate all $k$ documents. Since in most cases we can reasonably expect to not
cover all relevant keyphrases for $e$ in one or two documents, it is
imperative to keep the user engaged - requesting more documents, which in turn
can lead to better coverage. As soon as the user encounters a series of
inconsequential documents we can assume that she will quickly stop
requesting more documents. The detailed process is as follows:

\begin{enumerate}
	\item The user provides a set of names $\mathcal{N}$ and an (optional) initial
	description in the form of keyphrases $\mathcal{K}$. These keyphrases can
	be very few, maybe only one or two highly salient ones. If the name is not
	too ambiguous, and the correct entity is likely to show up frequently
	enough, giving initial keyphrases might not be necessary. In any case, the
	keyphrases are only needed to get the actual process started; after this,
	the user never has to actively provide keyphrases anymore.
	\item Candidate documents $\mathcal{D_{\text{cand}}} \subset \mathcal{D}$ are
	retrieved by querying $\mathcal{D}$ for all the strings in $\mathcal{N}$ (and $\mathcal{K}$).
	\item The user is shown the top ranked document with its EICs and has to make the following decisions: 
	\begin{itemize}
		\item The user either accepts or rejects the document, i.\,e.\ stating that the entity shown does or does not correspond to the one to be added.
		\item If the user accepts the document, all keyphrases $\mathcal{K}_d$ are mined from the document and presented to the user. The user can decide which keyphrases from $\mathcal{K}_d$ are added to $\mathcal{K}$. All rejected keyphrases are added to $\mathcal{K}^-$. If the document is rejected by the user,  $\mathcal{K}_d$ is added to $\mathcal{K}^-$. 
		
	\end{itemize}
	\item After keyphrase selection, the ranking is then re-computed based on the feedback from the user. The ranking is generated by one of
		the retrieval methods described in Section~\ref{sec:approach}. 
		 Note that ranking with small
		initial $\mathcal{K}$ can easily go wrong, which is exactly why user
		feedback is necessary.
	\item Once the user is finished, the entity is added to the KB using $\mathcal{K}$ as the description.
\end{enumerate}

When the entity is finally saved to the knowledge base, additional statistics
like co-occurrence counts between entities and keyphrases can be mined from
the accepted documents to, for e.\,g.\ compute keyphrase weights further improving the
disambiguation quality.

The most important part of this process, and the one that this paper will be focusing on, is the ranking of documents so that the number of unique keyphrases encountered is maximized while keeping the user engaged.

\section{Metrics}
\label{sec:metrics}

\textbf{Notations.} For a given query $q$ intended to add entity $e$, we denote the set of all relevant keyphrases as $\mathcal{K}_e$. Let $S \subseteq \mathcal{D}_{cand}$ be the set of all documents encountered by the user ($|S|=i$). $\mathcal{K}_{cand}$ is the set of all keyphrases mined from $\mathcal{D}_{cand}$ and let $\mathcal{K} \subseteq \mathcal{K}_{cand}$ be the set of keyphrases added by the user so far. $S_R \subseteq S$ denotes the set of all documents judged relevant by the user and $\overline{S_{R}}$ is the set of irrelevant documents. 

Since the key requirement for adding new entities is the suitability of the representation $\mathcal{K}$ for disambiguation, the natural metric to measure would be its \emph{disambiguation accuracy}. Unfortunately, this cannot be optimized or used to guide our algorithms, as there is no ground truth data to compute this. Because of this, we use two intrinsic measures conforming to the goals of our problem -- \emph{coverage} and \emph{engagement}.

\textbf{Coverage:} It is defined as the fraction of relevant keyphrases selected by the user from the EICs she has evaluated so far. The coverage at $i$ for the entity $e$ is given by: 
\begin{equation}
  Cov@i = \frac{\big| \mathcal{K} \cap \mathcal{K}_e \big|}{|\mathcal{K}_e|}
\end{equation}







\textbf{Engagement:} Measuring user engagement directly is a challenging endeavor. However it can be estimated indirectly by observing phenomena that can lead to increased engagement. According to~\cite{lalmas2014measuring}, user engagement can be estimated indirectly by novelty of information encountered by the user. 

To measure engagement in our scenario, we first assume that the user is more likely to abandon the task when she encounters a sequence of \emph{inconsequential} documents. A documents is called inconsequential if it does not add to $\mathcal{K}$ (i.e. no novel keyphrases for the user). The more documents of consequence (documents which add a new keyphrase) the user sees in a sequence the more engaged she is.  We denote $\mathcal{I}_S$ as the set of all \emph{maximal inconsequential sequences} of documents for $i$ documents encountered and the set of documents of consequence as $\mathcal{C}_S$. We define $Engagement@i$ as:
\begin{equation}
  Engagement@i = \frac{\sum_{\gamma \in \mathcal{I}_S}  \frac{1}{1+|\gamma|} + \sum_{\gamma \in \mathcal{C}_S} |\gamma|}{i}
\end{equation}





For ideal engagement $\mathcal{I}_S = \phi $ meaning all documents ordered in $S$ are consequential and $Engagement @i = 1.0$. 

For example, consider two lists $A = <+, -, +, -, +, ->$ and $B = <+, +, -, -, -, +>$ where $-$ denotes an inconsequential document and $+$ denotes a document of consequence. $A$ with inconsequential sequence lengths of $\{1,1,1\}$ ($engagement = 0.75$) is more engaging than $B$ with a single inconsequential sequence of length $3$ ($engagement = 0.54$) although it has the same number of inconsequential documents because the user is motivated each time she finds a document of consequence. Note that we differentiate between an irrelevant document and an inconsequential document, since a document might be relevant to the query, yet not add new keyphrases to $\mathcal{K}$.



\section{Approach}
\label{sec:approach}

\note{
\begin{enumerate}
	\item language model(b1): doesn't take feedback into account. feedback is good because user tells us which part of the context is relevant. we also see relevant keyphrases usually co-occur.
	\item some keyphrases are more relevant than others
	\item language model with active expansion (b2): takes feedback into account, not very robust since it could start with crap keyphrases
	\item lang model interleaved with expansion lang model: more robust but still covers the space  slowly
	\item add diversity into the mix to cover keyphrases faster
	\item interleave static entitiy diversification with expansions (a1) - robust but state not maintained -plateaus
	\item interleave static and expansion based ranking only when gradient is 0. also maintain the state of diversification(a2) - ent diversification and kp diversification can be used
	\item pseudo code needed for each approach
\end{enumerate}
}
An approach designed for the task of helping the user gather context for a new ambiguous long tail entity should possess the subsequent desirable properties following the goals described before (in Section~\ref{sub:problem}): 

\paragraph{Take user feedback into account} The user explicitly states which keyphrases are relevant. While this feedback would have no impact when trying to purely cover the keyphrase space, we find that \emph{relevant keyphrases} tend to co-occur with each other and can help guide the retrieval model towards covering the more relevant subset of the universal keyphrase space. This intends to minimize user effort to cover keyphrases by assessing a small number of documents.

\begin{figure}[b!]
  \centering
  \includegraphics[width=0.3\textwidth]
  {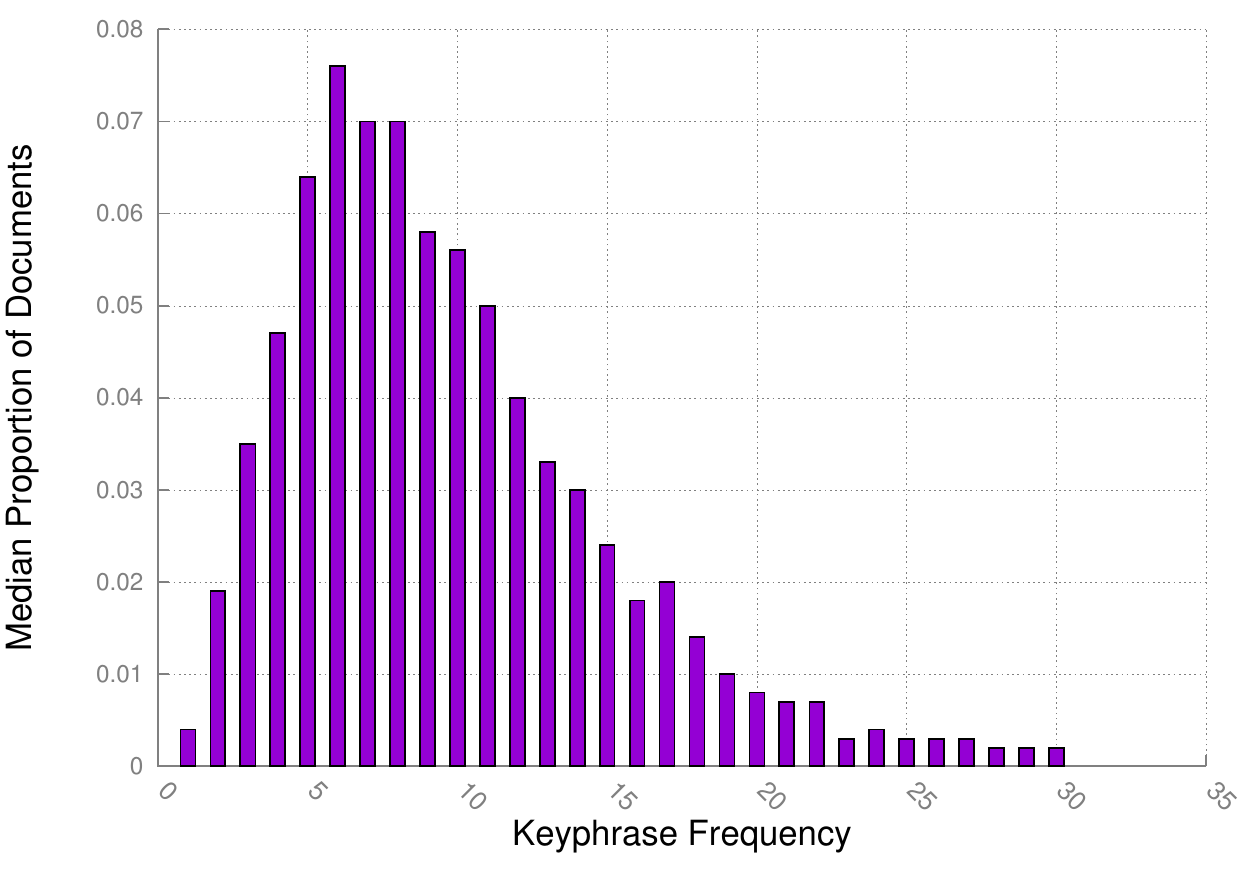}
  \caption{Keyphrase co-occurence distribution for all entities in the workload (c.f.~\ref{sub:workload}): X-axis is the \#co-occurring keyphrases in a doc. Y-axis ratio of documents which have this co-occurrence frequency.}
  \label{fig:kpdist}
\end{figure}


To test the hypothesis that relevant keyphrases co-occur, we selected documents tagged with entities with high confidence from our document collection (cf. Section~\ref{sub:workload}) and counted the number of relevant keyphrases in the vicinity of the entity mention. From Fig.~\ref{fig:kpdist} we can see that most documents contain 4 - 12 co-occuring keyphrases. We rarely find documents with just a single keyphrase.

\paragraph{Engage the user} The user controls the number of documents presented to her. It is highly likely that by presenting the user with a sequence irrelevant documents or documents covering no new keyphrases she becomes disillusioned with the task and abandons it. If she is unsatisfied early on, we may have little to no context for the entity the user is trying to add. The longer the user is engaged with the task, the more keyphrases are likely to be covered. 

\paragraph{Robustness} The resultant ranking of a desirable approach should be fault tolerant especially when taking feedback into account. Ideally the algorithm should guide the user away from irrelevant keyphrases in $\mathcal{K}_{cand}$ and towards the more relevant keyphrases.

In the remainder of this section we discuss possible approaches that have at least one of these properties before presenting our algorithm that takes all three properties into account.  

\subsection{Relevance and Feedback}
\label{sec:lm-feedback}

The naive approach to retrieving documents relevant to the input $q$ is to rank based on textual relevance of the entity name and keyphraes (for instance using language models, denoted by \lm{} in our experiments). While textual relevance may be a good indicator for documents containing the entity, the retrieval model is not geared towards helping the user find more new keyphrases. Given our observation that relevant keyphrases co-occur we can use retrieval as a means to find new keyphrases by expanding the query with keyphrases already marked relevant by the user. Query expansion produces a new disjunctive query with more terms which results in the likelihood of increased recall. 

We consider a modified version of the classical Rocchio's algorithm for incorporating relevance feedback to formulate the query expansions. Specifically, unlike classical scenarios where the terms for expansion must be intelligently mined from the feedback documents, we have explicit judgements from the user regarding important terms found in a document. Rocchio's algorithm forms a new query vector by adding the result of the difference between the vector of terms representing the relevant documents and the vector of terms representing the irrelevant documents to the original query. In our scenario we model the query and document as vectors of keyphrases in $\mathcal{K}_{cand}$. 

We use the \emph{normalized keyphrase frequency} of the keyphrase in the document as weights in the query vector.



Using query expansion we can now guide the selection of new keyphrases which occur in documents containing keyphrases previously marked relevant by the user. Since we show a single document at each stage, one strategy of triggering retrieval with the newly expanded query is to do so every time we encounter a relevant document. As the user finds new keyphrases, the query vector gets updated and the textual relevance of the document indicates that it contains keyphrases seen before and also likely to contain new keyphrases. In our experiments we use this approach as a baseline (\lmFeedback{}). While this approach takes user feedback into account, it is highly susceptible to \emph{specialization}. 



\textit{Specialization} occurs when the user encounters a relevant document with no new relevant keyphrases, i.e. all keyphrases mined from the EIC's in this document have already been added to $\mathcal{K}_{cand}$ or none of them are relevant to the entity.

If the user selects keyphrases pertaining to only one aspect of the entity then the query vector \emph{specializes} to retrieve very similar documents resulting in user disengagement.



\subsection{Diversification} 
\label{sub:diverse_document_ranking}

Since we are interested in optimizing \emph{relevant keyphrase coverage}, an 
intrinsic measure for our problem, search result diversification approaches like~\cite{agrawal2009diversifying} are a natural fit to our scenario.
Typical diversification approaches rely on the accurate modeling of underlying 
intents or aspects of a query. In our scenario, we naturally select the set of 
keyphrases $\mathcal{K}_{cand}$ mined from $\mathcal{D}_{cand}$ as the set of query aspects. 
Apart from optimizing coverage, diversification also addresses specialization of presented documents by actively seeking to discredit documents which contain already covered keyphrases. Formally, we are interested in selection of documents $S$ in a sequence such that maximizes coverage of $\mathcal{K}_{cand}$, i.e.,
$$
  argmax_{S} \,\,coverage(\mathcal{K}_{cand})\,\,, s.t. \,\,|S| \leq i.
$$

Akin to~\cite{agrawal2009diversifying}, we employ a greedy algorithm to approximate this NP-hard problem with a proven approximation guarantee of $(1-\frac{1}{e})$. It greedily chooses a document $d$ at each iteration which has the highest marginal coverage with respect to $S$. The choice of the greedy solution is also a natural fit to our scenario in which we are unsure when the user leaves. Hence, we would want to maximize the coverage at each step of the addition process. 

However a major drawback arises while using such an approach for ambiguous entities where query is likely to be underspecified. This means that not all the aspects from $\mathcal{K}_{cand}$ are relevant to $e$. This might result in retrieval of documents which do not cover the \emph{relevant} keyphrase space and hence leads to \emph{concept drift}.

\textit{Concept Drift} in our scenario is said to occur when the user encounters an irrelevant document, i.e. a document unrelated to $e$. 


\textbf{Incorporating Feedback: } Without taking feedback into account and given the ambiguity of the entity being added, we may encounter concept drift by initially selecting documents that cover irrelevant aspects of the entity. To account for concept drift we take consider user feedback in order to refine $\mathcal{K}_{cand}$ by altering $\mathcal{D}_{cand}$ with a new expanded query similar to the previous subsection. 

One major modification is necessary to incorporate user feedback while still diversifying result lists. That is, the state of the selected keyphrases should be maintained and utilized while re-ranking the retrieved documents ensuring diversity. Consequently, after every expansion we diversify results keeping in mind that $\mathcal{K}$ has already been added. We refer to the vanilla diversification approach over the keyphrase space as \divKp{} and diversification with feedback as \divKpFeedback{}.

\subsection{Interleaving Result Lists}
\label{sec:interleaving}
Named Entity Disambiguation (NED) systems struggle when disambiguating mentions of entities which have high context overlap. For such entities we need more \textit{discriminative} keyphrases to improve the disambiguation accuracy. \divKpFeedback{} is designed to overcome both specialization and concept drift but it is a safety-first approach. For example, consider the user is trying to add the entity \textit{Perseus}, the constellation, not the mythical Greek hero after whom it is named. Both entities share many keyphrases since they are related. There are two major aspects to this entity: first, the origin of the name and connection to other constellations named after Greek heroes (\texttt{ptolemy}, \texttt{greece}, \texttt{andromeda}) and second is astrological context (\texttt{galactic plane}, \texttt{milky way}). Let us assume that the initial documents presented to the user contain relevant but \textit{non-discriminative} keyphrases (keyphrases from the first aspect). Utilizing \divKpFeedback{}, firstly there is no guarantee that discriminative keyphrases will co-occur with the non-discriminative ones selected by the user. This will likely lead to a description void of discriminative keyphrases which are crucial for improved disambiguation accuracy. Second, if all relevant keyphrases for this aspect have been covered, the user will have to evaluate inconsequential documents.

To account for this, we first assume that $\mathcal{D}_{cand}$ produced by a retrieval model like \lm{} or \divKp{} for the initial query contains keyphrases from all aspects of the entity. We propose interleaving results from the feedback-based approaches producing \emph{dynamic lists} (\lmFeedback{} or \divKpFeedback{}) with the baseline approaches which do not consider feedback or produce \emph{static lists} (\lm{} or \divKp{}). Specifically, given two lists, the \emph{static} $A = <a_1, a_2, \ldots>$ and the \emph{dynamic} $B = <b_1, \ldots>$, we can present results in an interleaved manner to the user $I = <a_1, b_1, a_2, \ldots>$. However, the dynamic list updates continuously due to query expansions based on the positive/negative feedback from the user. In the Perseus example, the dynamic list allows the user to add non-discriminative keyphrases while the static list can be used to find new discriminative keyphrases.

\begin{figure}[t!]
  \centering
  \includegraphics[width=0.4\textwidth]
  {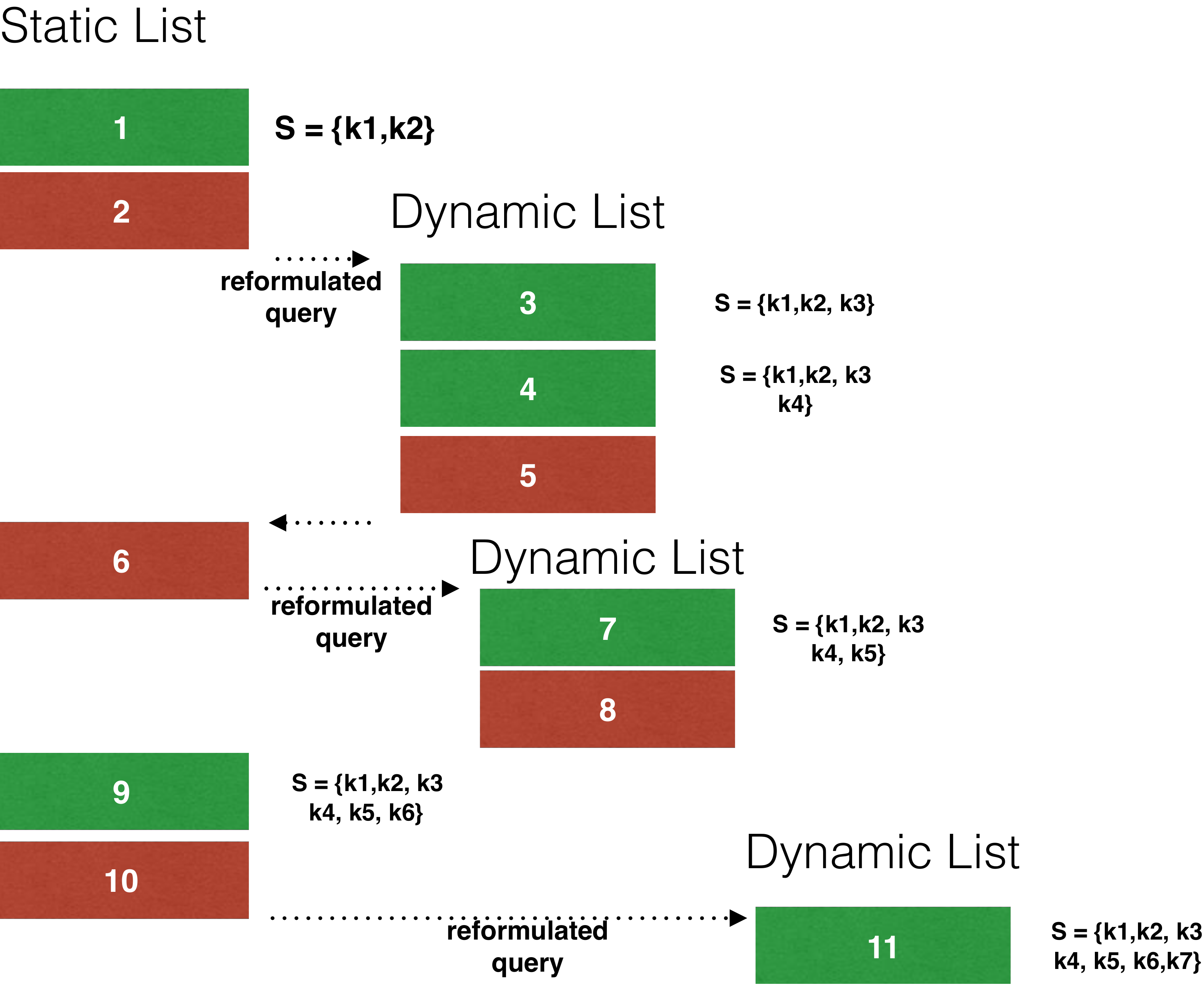}
  \caption{Interleaving result lists. Green rectangles represent relevant documents, red rectangles represent non-relevant documents.}
  \label{fig:interleaving}
\end{figure}

A naive procedure for switching between the lists could be executed in a predetermined manner, i.e., alternate between $A$ and $B$ producing $I = <a_1, b_1, a_2, b_2 \ldots>$. Although, such a procedure ensures robustness, it might not engage the user actively. Especially in scenarios when the dynamic list correctly converges to the relevant keyphrase space switching to the static list might be undesirable. Instead we actively keep track of the user assessments for each list and dynamically decide the next document based on the last successful assessments. In particular, we defer switching until we encounter an inconsequential document, i.e., a document where the user does not add a new keyphrase. An illustration of this is shown in Figure~\ref{fig:interleaving} where top-5 of the 11 documents shown to the user are chosen from the static list. Note that whenever we encounter an inconsequential document in the static list the query vector is recomputed given the state of $S$, a new ranked list of documents is retrieved, and the top document of this list is presented to the user. 




We refer to the interleaved approaches as \textsc{I(.,.)} where the first argument is the static list(\lm{} or \divKp{}), and the second argument is the approach with feedback (\lmFeedback{} or \divKpFeedback{}), e.g., \interleavelmlm{}.


\subsection{Further Design Decisions}
\label{sec:final-approach}

In designing an effective approach with the above ingredients, we are left with two design decisions: \emph{Which aspect space should we diversify over?} and \emph{Which approaches do we use to interleave?}

\textbf{Diversifying over entities} Although using the keyphrases to model query aspects might seem natural, there are certain disadvantages to it. First, the keyphrase space is large. Consequently, the diversification algorithm always has a choice to present a document with new keyphrases due to its inherent nature to explore uncovered keyphrases. While this is desirable on one hand, documents which contain already covered keyphrases are seldom preferred. Since we know that relevant keyphrases tend to co-occur, this in fact is detrimental to the coverage of the relevant keyphrase space. Secondly, it is hard to canonicalize keyphrases. A key component in the diversification algorithm is spotting if a pair of aspects are indeed the same. While this is easy for  words there is no straightforward way in which two keyphrases (differently phrased) can be identified accurately. For example, \texttt{Languages of Nigeria} and \texttt{Nigerian Languages}.

We identify that the \emph{canonicalized entities} on the other hand are better aspect representations than keyphrases. Entities are much smaller in number and are easily identifiable. More importantly, NED systems routinely employ joint inferencing over entities present in the document to better disambiguate the mention in question. We refer to the diversification approach using the entity space as \divEnt{} and its feedback counterpart as \divEntFeedback{}.

\textbf{Interleaving with diversified lists} Since we have to choose an approach with feedback (dynamic) and another without feedback (static) we have nine choices from \lm{}, \divEnt{}, \divKp{} and their feedback-based counterparts. However, we have already seen that entity-based diversification approaches possess an inherent advantage over keyphrase-based diversification. Secondly, as explained in Section~\ref{sec:lm-feedback}, \lmFeedback{} is prone to specialization. Although interleaving it with \lm{} (\,\interleavelmlm{}\,) renders it resilient to specialization, it might cause concept drift because if a partially-relevant yet non-salient keyphrase $p$ is selected early on the expansions would tend to specialize to documents containing $p$.

Thus we choose \divEntFeedback{} for generating the dynamic ranking and \lm{} or \divEnt{} as the static ranking (denoted as \interleaveLmDivEntFeedback{} and \interleavedivEntEntFeedback{}). In the experiments we additionally consider \interleavedivKpKpFeedback.

\section{Experimental Setup}
\label{sec:setup}

\note{
\begin{enumerate}
	\item explain the kb used, the corpus and the disambiguation tool
	\item how did we generate the workload - mention, 3 keyphrases taken from kb
	\item workload stats 
	\item metrics - coverage, weighted coverage (precision)
	\item mention baselines used
	\item keyphrase mining technique
	\item simulate the user - high precision annotations , kps from the kb
	\item removed wikipedia, foreign language documents and duplicates
\end{enumerate}
}
In this section we describe the experimental setup we used to empirically evaluate the performance of our approach and other baselines. We conducted a small scale experiment with real users and a large scale user simulation.

\subsection{Document Collection and Knowledge Base} 
\label{sub:document_collection}

We choose AIDA~\cite{Hoffart:2011a} as the NED system for our experiments. AIDA is a state-of-the-art NED system that performs joint disambiguation based on context overlap with keyphrases. It links mentions to corresponding entities in YAGO. AIDA represents the context of an entity by a ranked list of keyphrases mined from a Wikipedia dump (from 2014 in our experiments). We denote the entity representations -- canonicalized entity along with its associated keyphrases, maintained in AIDA as \texttt{EntityKB}. As an external text corpus to find relevant keyphrases for entities, we used the ClueWeb09 corpus which consists of approximately 50 million web pages from a crawl conducted in 2009. We restrict ourselves to English documents and remove all duplicates.

\subsection{Users} 
\label{sub:user_simulation}

There are two kinds of relevance judgments that the user must provide. The first is binary relevance on a document level - \emph{Is the document relevant to the entity I am trying to add?} and the second is binary relevance for the keyphrases found in a relevant document - \emph{Is $k$ a relevant keyphrase for the entity I am trying to add?}
 
Given the interactive nature of our scenario, we cannot rely on traditional IR evaluation techniques like pooling. Even the slightest change in user judgments could mean encountering new documents that have yet to be judged. Assuming these documents to be irrelevant will introduce significant noise in the results.

We opt to simulate the user instead by indirectly gathering relevance judgments for all documents. To identify documents relevant to a particular entity we use the annotations from the FACC1 dataset released by Google~\cite{gabrilovich2013facc1}. FACC1 consists of high precision automatically extracted entity mentions that are linkable to the Freebase knowledge base from all documents in ClueWeb09. We assume a document to be judged relevant for an entity $e$ if it has been tagged with $e$ according to FACC1. For the keyphrase level judgments, we declare a keyphrase relevant if it also occurs in \texttt{EntityKB}. In this way we focus the evaluation on relevance judgments for documents rather than keyphrases which are assumed to be identified correctly by the user.

Another key aspect of simulating the user is to determine the query the user will use to add an entity to a knowledge base. In an ideal world we would have a query log or users suggesting entities that do not exist in the knowledge base. However the former is non-existent and the latter is an exhaustive and expensive procedure. Instead we generate a workload by removing existing long tail entities in the knowledge base which have a reasonable number of keyphrases present in the collection. 


For completeness and to examine the effect of real user judgments, we also conducted a small scale experiment with students. Refer to Section~\ref{sub:user_study} for the user study.

\subsection{Query Workload and Ground Truth}
\label{sub:workload}
We can reasonably expect the user to use a mention and a few supporting keyphrases that describe the entity as the initial query. For example if the user is trying to add the entity \textbf{Chris Jericho} (a professional wrestler), a reasonable query would likely consist of an ambiguous mention like \textsf{jericho} along with keyphrases like \texttt{pro wrestling}, \texttt{edge} and \texttt{the rock}. To identify similar queries and semi-automatically generate a workload we first define certain criteria: 

\begin{enumerate}
	\item An entity is assumed to be long tail if it does not occur very frequently in the collection at hand. We also want queries which have a reasonable number of relevant documents. Hence we set an upper bound on the document frequency of the entity to 2000.
	\item A long tail entity is assumed ambiguous if a popular mention of this entity can also be linked to other entities in YAGO. A mention is considered popular for an entity if it has a high prior probability of being linked to this entity directly. For example \textsf{jericho} is a popular mention for \textbf{Chris Jericho} but it is ambiguous since \textsf{jericho} can also be linked to \textbf{Jericho City}. We can compute this from \texttt{EntityKB}. 
	\item We only consider long tail ambiguous entities with reasonable coverage of its' keyphrase set given the collection. We set the lower bound of the number of keyphrases found in the collection to 50.
\end{enumerate}

Based on these requirements we generated a workload of 50 entities that we subsequently removed from \texttt{EntityKB}. The supporting terms for each entity query were selected from its' top 3 keyphrases in AIDA irrespective of its' presence in the collection. The set of all keyphrases found in the collection (which is a subset of all keyphrases mined from Wikipedia for the entity) forms the ground truth keyphrases. On average we found 200 relevant keyphrases per query in our workload. Note that to accurately simulate the scenario we also removed all Wikipedia pages from the collection since in our scenario the Wikipedia page for the entity does not exist at the time.

\subsection{Disambiguation Accuracy Groundtruth} 
\label{sub:disambiguation_accuracy_groundtruth}

For each entity in our workload, we randomly select 100 documents tagged with it as the disambiguation ground truth. FACC1 contains mentions tagged with entities from Freebase with high confidence. Even though our \texttt{EntityKB} is derived from YAGO, both knowledge bases have the same substrate -- Wikipedia. On average each document has 2-3 mentions of the entity. The disambiguation accuracy is calculated as the percentage of mentions in the ground truth documents correctly tagged with entity using the output entity representation computed by an entity-addition approach. In our case, each output entity representation is added to \texttt{EntityKB} and we use AIDA to disambiguate these mentions. Disambiguation accuracy is computed for all entity representations at varying retrieval depths for all entities in our workload.

\subsection{Baselines} 
\label{sub:baselines}

We consider 3 distinct categories of approaches. 

\textbf{Language Models.} The first category is all retrieval models based on \emph{textual relevance}. We use a statistical language model with Dirichlet smoothing ($\mu=1000$) called \lm{} as the baseline. This baseline represents pure textual relevance without incorporating any user feedback. Next we consider the case where we initially rank by \lm{} but incorporate feedback by actively expanding the query (trigger retrieval each time the user encounters a relevant document) using the Rocchio algorithm -- \lmFeedback. Next we consider our more robust variant \interleavelmlm{} that interleaves \lm{} and \lmFeedback{} (described in Section \ref{sec:lm-feedback}). 

\textbf{Keyphrase based.} The second category consists of retrieval models based on diversification using \emph{keyphrases} as the \emph{aspect space}. Keyphrases can be automatically mined in several ways. For example, regular 
expressions over part-of-speech patterns can be used to harvest keyphrase candidates. These patterns would serve as a filter to include useful phrases. 

In practice, noun phrases that include proper nouns (i.e. names or parts of names) and technical terms have been shown to be useful~\cite{Hoffart:2014hp}. We consider the standard greedy approach to diversification suggested by \cite{agrawal2009diversifying} and use keyphrases as aspects for diversification.  We call this baseline \divKp. Akin to \lmFeedback{} and \interleavelmlm{} we also have \divKpFeedback{} and \interleavedivKpKpFeedback. Additionally we also consider \interleaveLmDivKpFeedback.

\textbf{Entity based.} Finally we consider diversification using \emph{entities} as the aspect space. We use the annotations provided by FACC1 as entity aspects for diversification. Similar to the keyphrase category, the entity category consists of 3 approaches: \divEnt, \divEntFeedback{} and \interleavedivEntEntFeedback. We also consider interleaving \lm{} with \divEntFeedback{} -- \interleaveLmDivEntFeedback. For all interleaving approaches we interleave the top 20 documents of the static ranking with the dynamic ranking.

Lastly to put things in perspective we compute an ideal ranking for each query. We assume that the ideal ranking covers the maximum number of unique relevant keyphrases at each step. \ideal{} represents the retrieval model that greedily optimizes for maximum set cover of only the relevant keyphrases from the ground truth. 



\begin{table}[t]
  \footnotesize
  \centering
   \scalebox{0.85}{
  \begin{tabular}{@{}llcccc@{}}\toprule

    \multicolumn{2}{l}{} & 5 & 10 & 15 & 20\\
    \midrule

    \multicolumn{2}{l}{\lm} &10.44\%&17.06\%&18.51\%&22.00\%\\
    \midrule
    \multicolumn{2}{l}{\lmFeedback} &9.16\%&18.91\%&19.25\%&22.17\% \\
    \multicolumn{2}{l}{\interleavelmlm} &10.41\%& 16.21\%& 17.75\%& 20.52\%\\
    \midrule
    \multicolumn{2}{l}{\divKp} &10.84\%&14.97\%&16.21\%&16.82\%\\
    \multicolumn{2}{l}{\divKpFeedback}& 10.64\%&14.72\%&17.79\%&18.83\%\\
    \multicolumn{2}{l}{\interleavedivKpKpFeedback} &9.95\%&14.72\%&16.94\%&18.81\%\\
    \multicolumn{2}{l}{\interleaveLmDivKpFeedback} &12.40\%&18.09\%&20.30\%&21.15\%\\
    \midrule
    \multicolumn{2}{l}{\divEnt}&14.24\%&21.56\%&23.53\%&24.51\%\\
    \multicolumn{2}{l}{\divEntFeedback}& 13.18\%&21.14\%&24.40\%&28.01\%\\
    \multicolumn{2}{l}{\interleavedivEntEntFeedback}& 12.34\%&21.29\%&24.55\%&27.76\%\\
    \multicolumn{2}{l}{\interleaveLmDivEntFeedback} & 15.06\%&23.88\%&27.07\%&29.78\%\\
    \midrule
    \multicolumn{2}{l}{\ideal} &15.96\%&27.28\%&32.56\%&36.56\%\\

    \bottomrule
  \end{tabular}}
  \caption{Disambiguation accuracy for all queries in the workload at $k=5,10,15,20$.}
  \label{tab:acc_all}
  \vspace{-5pt}
\end{table}

\begin{table}[t]
  \footnotesize
  \centering
   \scalebox{0.85}{
  \begin{tabular}{@{}llcccc@{}}
  \toprule
    \multicolumn{2}{l}{} & 5 & 10 & 15 & 20\\
  	\midrule

    \multicolumn{2}{l}{\lm} &16.17\%&26.37\%&28.65\%&34.00\%\\
    \midrule
    \multicolumn{2}{l}{\lmFeedback}& 14.18\%&29.27\%&29.80\%&34.31\%\\
    \multicolumn{2}{l}{\interleavelmlm}& 15.96\%& 16.21\%& 17.75\%&20.52\%\\
    \midrule
    \multicolumn{2}{l}{\divKp} &16.79\%&23.19\%&25.11\%&26.03\%\\
    \multicolumn{2}{l}{\divKpFeedback}& 16.47\%&22.79\%&27.54\%&29.16\%\\
    \multicolumn{2}{l}{\interleavedivKpKpFeedback} &15.41\%&22.79\%&26.24\%&29.11\%\\
    \multicolumn{2}{l}{\interleaveLmDivKpFeedback} &19.44\%&28.35\%&31.79\%&33.10\%\\
    \midrule
    \multicolumn{2}{l}{\divEnt}&22.05\%&33.34\%&36.16\%&37.55\% \\
    \multicolumn{2}{l}{\divEntFeedback} &20.42\%&32.54\%&37.79\%&42.95\%\\
    \multicolumn{2}{l}{\interleavedivEntEntFeedback} &19.16\%&33.00\%&37.93\%&42.62\%\\
    \multicolumn{2}{l}{\interleaveLmDivEntFeedback}& 23.89\%& 37.85\%&42.45\%& 46.86\%\\
    \midrule
    \multicolumn{2}{l}{\textsc{IDEAL}} &24.42\%&41.68\%&49.79\%&55.92\% \\

    \bottomrule
  \end{tabular}}
  \caption{Disambiguation accuracy for the subset of queries which have low context overlap with corresponding existing ambiguous entities in the $KB$  at $k=5,10,15,20$.}
  \label{tab:acc_low_context_overlap}
  \vspace{-5pt}
\end{table}

\begin{figure*}[ht!]
\begin{subfigure}{.5\textwidth}
  \centering
  \includegraphics[width=.8\linewidth]{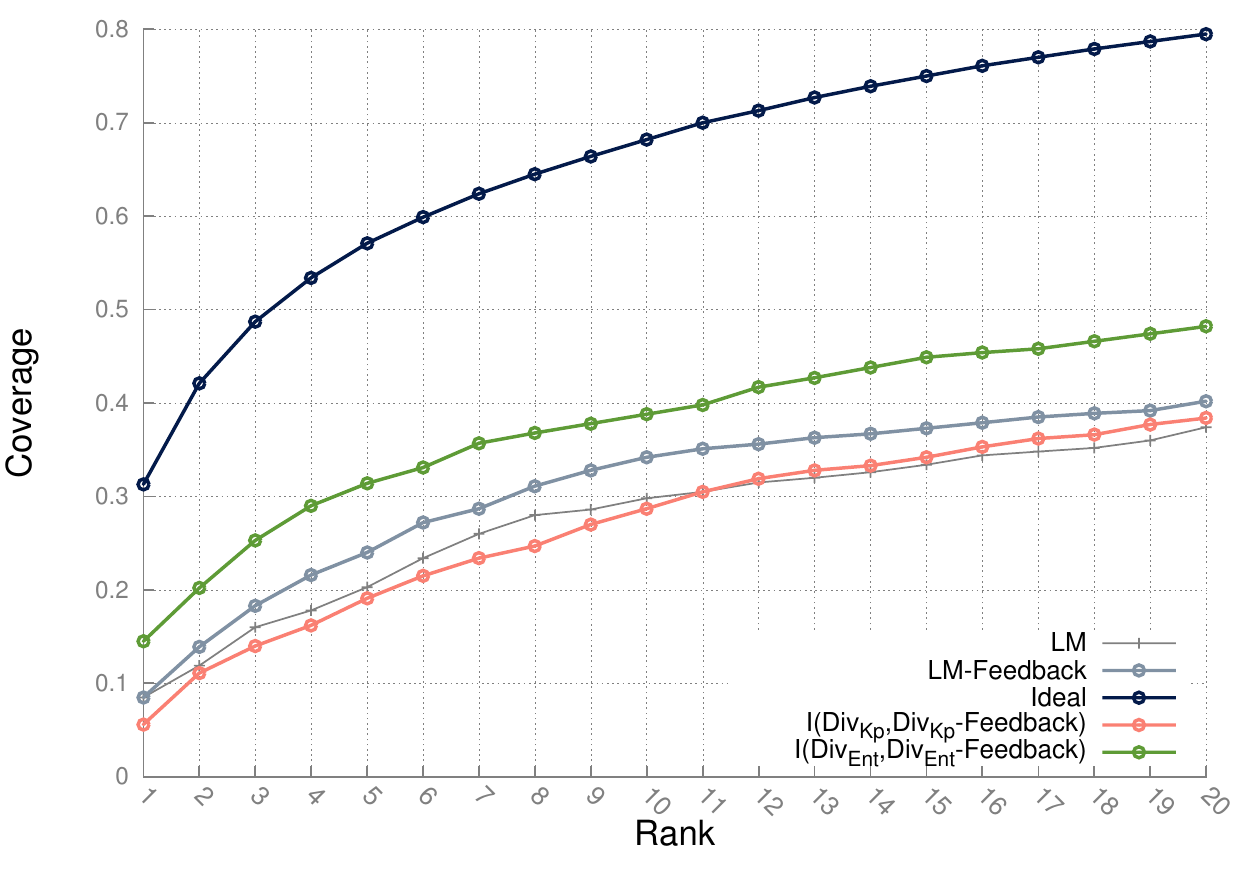}
  \caption{Top approaches in each category and the Ideal ranking.}
  \label{fig:sfig_best}
\end{subfigure}%
\begin{subfigure}{.5\textwidth}
  \centering
  \includegraphics[width=.8\linewidth]{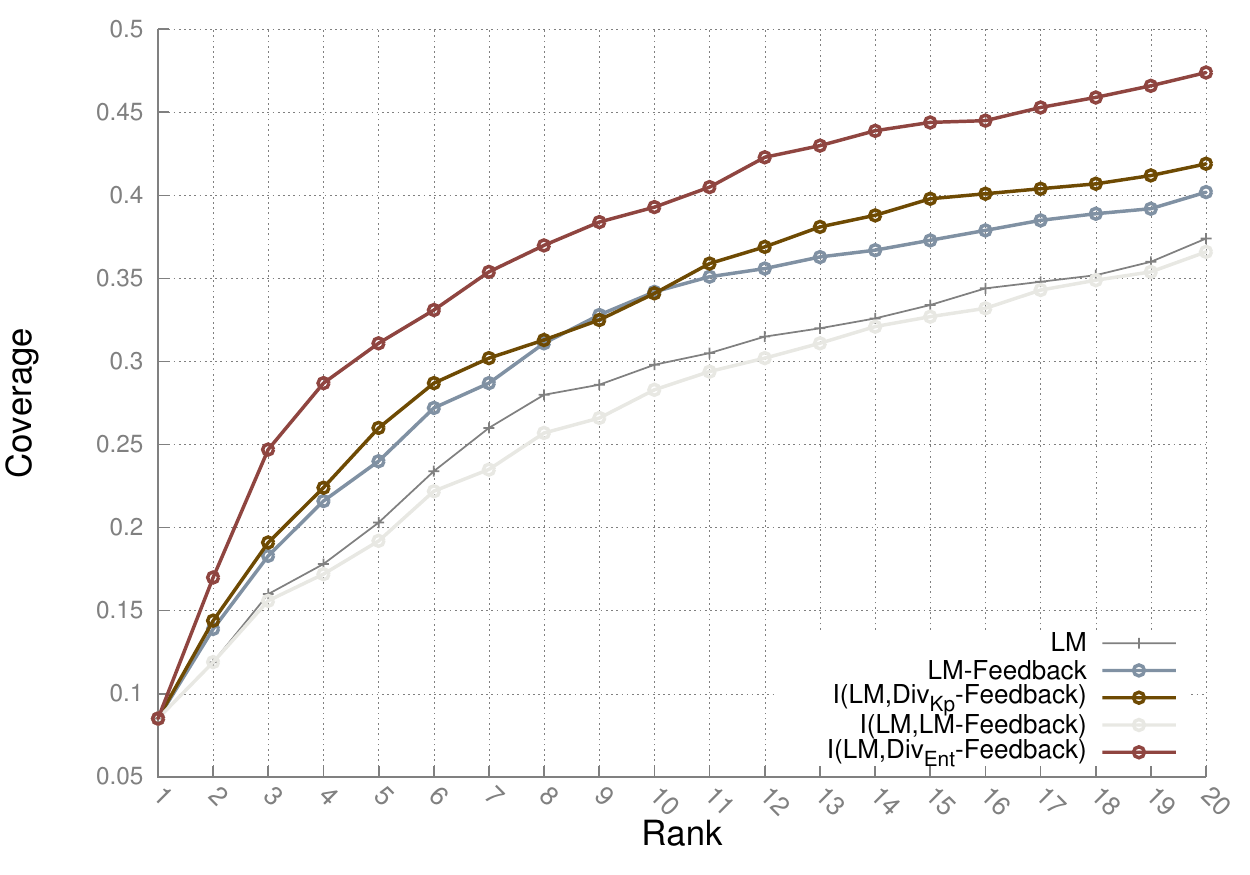}
  \caption{Language Model based approaches (Effect of rank 1).}
  \label{fig:sfig_lm}
\end{subfigure}
\begin{subfigure}{.5\textwidth}
  \centering
  \includegraphics[width=.8\linewidth]{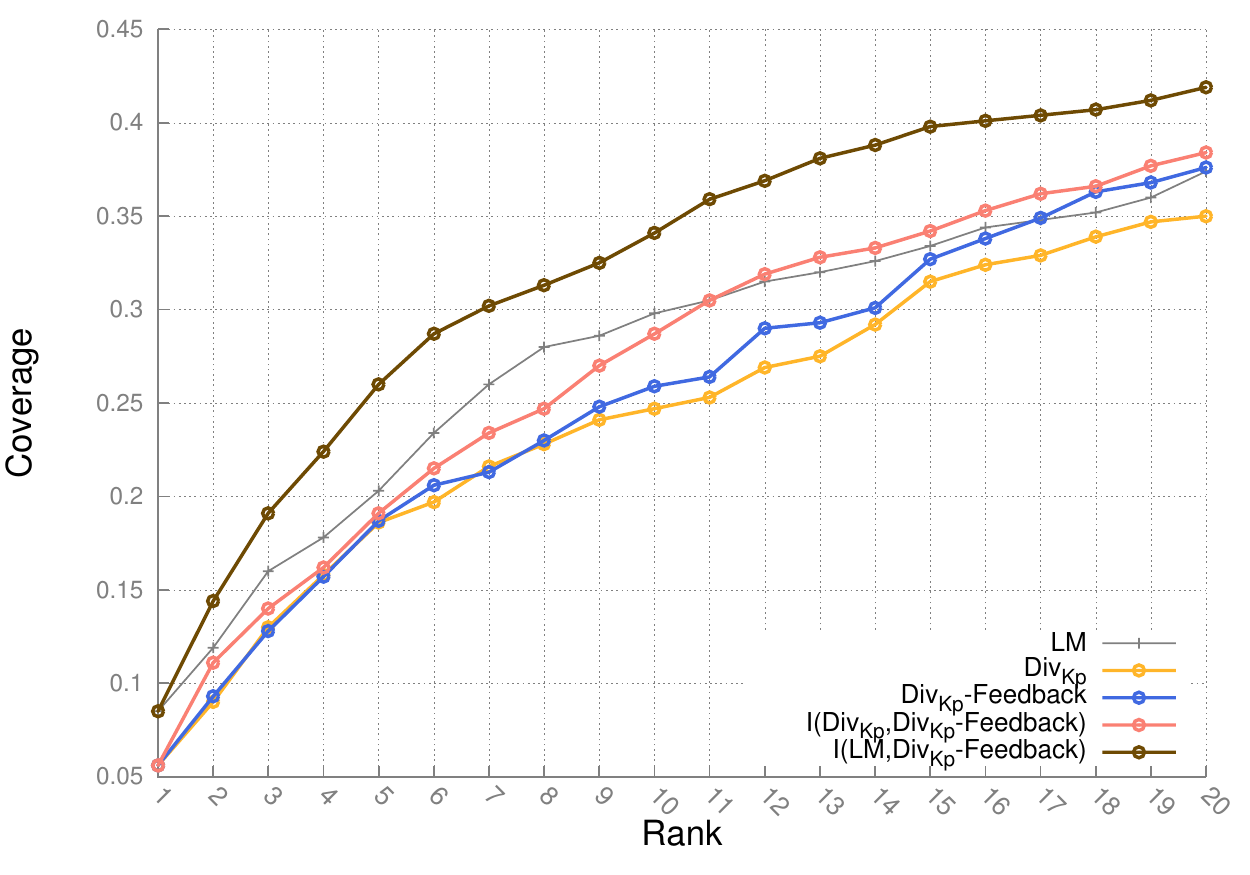}
  \caption{Keyphrase Diversification based approaches.}
  \label{fig:sfig_kp}
\end{subfigure}%
\begin{subfigure}{.5\textwidth}
  \centering
  \includegraphics[width=.8\linewidth]{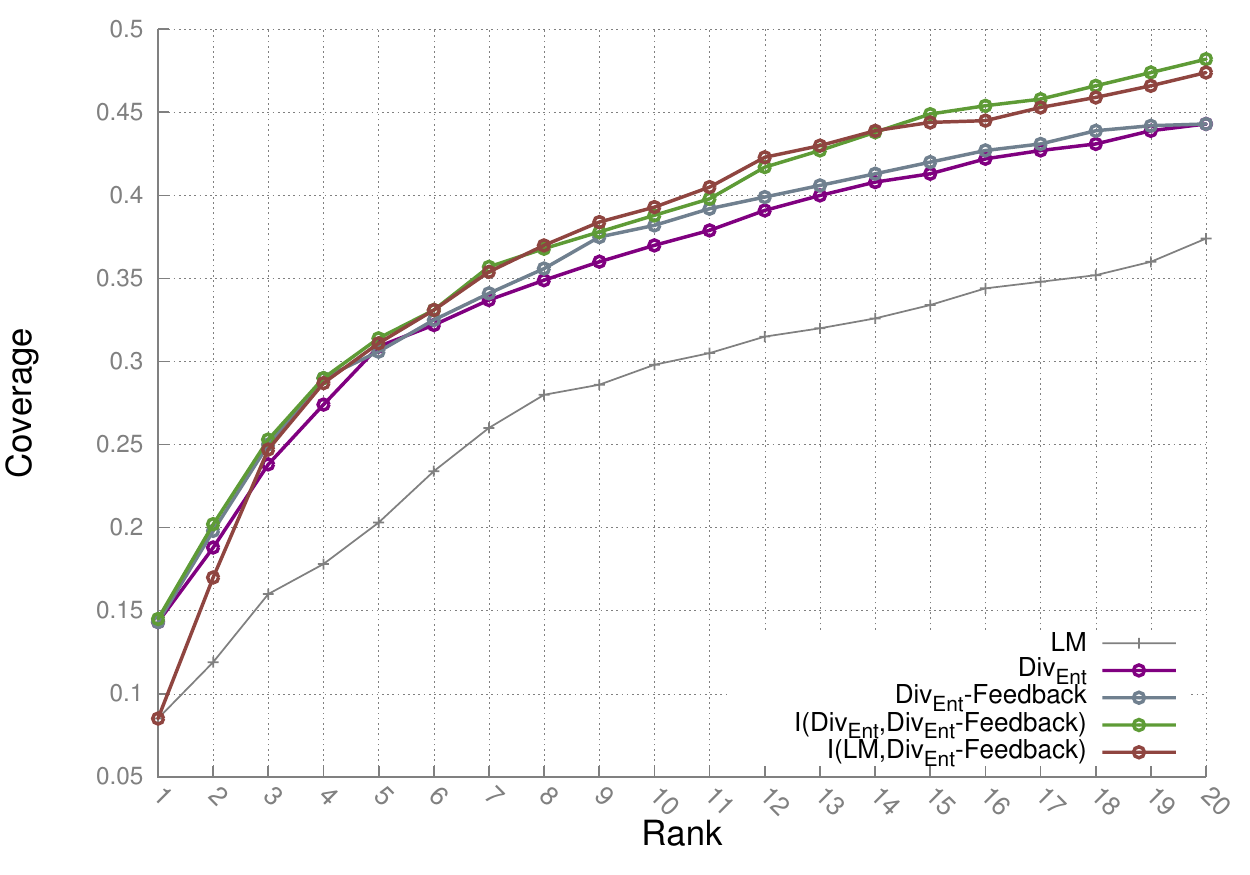}
  \caption{Entity Diversification Approaches.}
  \label{fig:sfig_ent}
\end{subfigure}
\caption{Keyphrase Coverage vs. Rank: The plots show the fraction of keyphrase coverage against the number of documents the user requests ($k=1$ to $20$). }
\label{fig:fig_coverage}
\vspace{-5pt}
\end{figure*}

\section{Results} 
\label{sec:results}

In this section we empirically determine the best approach for the task of entity in context addition using both extrinsic and intrinsic measures to study the effectiveness of the entity-addition approaches. We first present results from the user simulation over the entire query workload.


\subsection{Extrinsic Measures} 
\label{sub:extrinsic_measures}

The extrinsic measure of choice for our task is disambiguation accuracy. Table \ref{tab:acc_all} shows the performance of all approaches for all queries in the workload. At first we notice that entity based approaches are the best. \interleaveLmDivEntFeedback{} achieves the highest disambiguation accuracy at all ranks. Entity based approaches also outperform \lm{} based approaches which in turn, rather surprisingly, outperform most keyphrase diversification based methods. It shows that identifying the relevant keyphrase space for ambiguous long tail entities is a difficult task. Diversifying over non-canonicalized keyphrases is detrimental to disambiguation accuracy. Consequently, using canonicalized named entities with the same retrieval model improves disambiguation accuracy considerably (See \divEnt{} and \divKp) at all ranks. 

Next, we consider the effect of user feedback. There are two ways of using feedback: either actively or by interleaving. \lm{} is slightly improved by incorporating user feedback actively whereas interleaving is harmful. In fact, \interleavelmlm{} has the lowest disambiguation accuracy overall. This is because the lack of diversity leads to a specialized representation of the context. 

When considering the diversification approaches we find that user feedback consistently improves disambiguation accuracy but only considerably after $k=15$. This shows that once the user has provided sufficient feedback, the query vector is more stable and  more relevant documents are retrieved. The use of interleaving approaches where both rankings are diversified (\,\interleavedivEntEntFeedback{} and \interleavedivKpKpFeedback{}\,) does not improve the disambiguation accuracy in our scenario. However \interleaveLmDivKpFeedback{} and \interleaveLmDivEntFeedback{} are considerably better than \divKpFeedback{} and \interleavedivEntEntFeedback, indicating that interleaving is highly sensitive to the nature of the chosen retrieval models. By selecting two contrasting approaches (textual relevance and diversification) we effectively capture a better representation of the relevant keyphrases. While diversification helps to find new keyphrases, interleaving with \lm{} is able better combat concept drift as compared to a diversified ranking.

Notice that though the disambiguation accuracy values in Table \ref{tab:acc_all} seem relatively low, \textsc{Ideal} achieves at most a disambiguation accuracy of $36.5\%$. The overall accuracy is also low partly due to ambiguous entities with mentions that already have popular entities with very similar context which they can link to. For instance, \textbf{Donatella Versace} (the fashion designer) has high overlap with the entity \textbf{Versace} (the company she owns) for the mention \textsf{versace}. Table \ref{tab:acc_low_context_overlap} shows the results for the disambiguation accuracy experiments on a subset of the query workload consisting of queries with low context overlap with existing popular entities with the same mention. Note that the trends remain the same in both experiments while the disambiguation accuracy is improved across all approaches.

\begin{table}[t]
  \footnotesize
  \centering
   \scalebox{0.85}{
  \begin{tabular}{@{}llcccc@{}}
  \toprule
    \multicolumn{2}{l}{Win/Loss @} & 5 & 10 & 15 & 20\\
    \midrule

    \multicolumn{2}{l}{\lmFeedback}& 25/8&33/10&29/15&33/13\\
    \multicolumn{2}{l}{\interleavelmlm} &6/18&10/20&14/24&12/28\\
    \midrule
    \multicolumn{2}{l}{\divKp} &22/23&20/28&22/27&21/28\\
    \multicolumn{2}{l}{\divKpFeedback}&19/26&18/30&23/26&25/24\\
    \multicolumn{2}{l}{\interleavedivKpKpFeedback} &23/24&22/28&24/26&26/24\\
    \multicolumn{2}{l}{\interleaveLmDivKpFeedback} &24/19&31/17&36/11&31/14\\
    \midrule
    \multicolumn{2}{l}{\divEnt}&33/16&34/15&34/15&35/14\\
    \multicolumn{2}{l}{\divEntFeedback}&35/14&37/12&39/10&32/17\\
    \multicolumn{2}{l}{\interleavedivEntEntFeedback}&37/12&35/14&37/10&37/10\\
    \multicolumn{2}{l}{\interleaveLmDivEntFeedback}& 37/8&38/10&43/6&43/5\\

    \bottomrule
  \end{tabular}}
  \caption{Win/Loss w.r.t \lm{}in coverage at $k=5,10,15,20$.}
  \label{tab:winloss}
\end{table}

\subsection{Intrinsic Measures} 
\label{sub:intrinsic_measures}

Measures like coverage and engagement help us intrinsically understand which approach achieves better performance. In our approach and the other carefully considered baselines, we use keyphrase coverage as a proxy for high disambiguation accuracy. Figure \ref{fig:sfig_best} illustrates the coverage of the top baselines in each category from rank 1 to 20. In the same graph we also plot the coverage at rank $k$ for the ideal ranking (\ideal). 

\begin{figure}
\centering
  \includegraphics[width=0.4\textwidth]{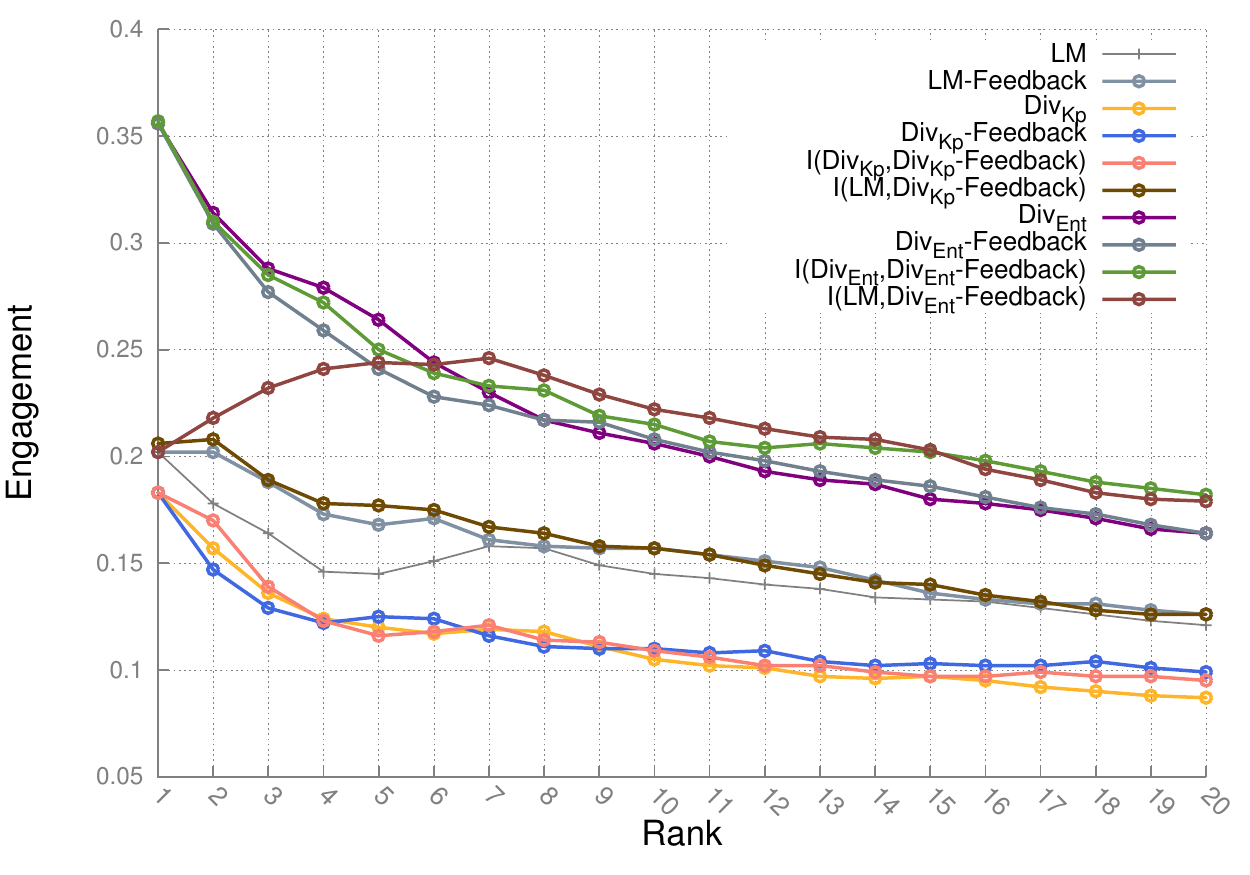}
\caption{Engagement}
\label{fig:eng_index}
\end{figure}

We see that given our setup, we can expect close to 80\% coverage with 20 documents. As expected we find that growth in keyphrase coverage is high in the beginning and stabilizes as $k$ increases. Once again the entity diversification based approaches perform significantly better\footnote{statistically significant difference between \lm{} and the top entity and keyphrase diversification approaches for p<0.05 using a paired t-test} than the other approaches in terms of coverage and engagement (see Figure \ref{fig:eng_index}). We also find that coverage is directly proportional to disambiguation accuracy in many cases. \interleavedivEntEntFeedback{} and \interleaveLmDivEntFeedback{} achieve close to 50\% coverage and the highest engagement at rank 20 while \interleaveLmDivEntFeedback{} also achieves the highest disambiguation accuracy.

Not surprisingly \lm{} is better than \divKp{} in terms of coverage and significantly better in engagement. The inability to canonicalize keyphrases leads to a large and noisy keyphrase space to diversify over (refer Section~\ref{sec:final-approach}). As a result the algorithm may easily drift causing the user to encounter many irrelevant documents. Figure \ref{fig:sfig_kp} illustrates the various keyphrase diversification approaches. Only \interleaveLmDivKpFeedback{} achieves higher coverage and better engagement than \lm{} at all $k>1$. Akin to the entity based approaches, we also see that interleaving in keyphrases improves coverage considerably after rank 5 (see \divKpFeedback{} vs \divKp).

Similar to the previous section we find that \lmFeedback{} is better than \lm{} but the difference in coverage is considerably higher when compared to the difference in disambiguation accuracy. Using keyphrases selected by the user to expand the query leads to significant improvement in coverage at all $k$, confirming the existence of co-occurring relevant keyphrases. Surprisingly we find that \lm{} suffers a dip in engagement until $k=6$. This dip is caused by the lack of new keyphrases in the top textually relevant documents. On the other hand, query expansion based on user feedback helps in finding textually relevant documents with new keyphrases, keeping the user engaged. 

In Figure \ref{fig:sfig_ent} we observe the coverage of all entity based diversification approaches. The coverage of all approaches is significantly better than \lm{} and \lmFeedback. Interleaving is effective at consistently finding new keyphrases whereas the other approaches tend to slow down at the higher ranks. Although \interleavedivEntEntFeedback{} and \interleaveLmDivEntFeedback{} achieve similar coverage, \interleaveLmDivEntFeedback{} obtains higher disambiguation accuracy. This indicates that both the quality and quantity(coverage) of keyphrases found plays a vital role in improving disambiguation accuracy. In other words, although the former has a larger coverage of relevant keyphrases the latter covers the more important relevant keyphrases.

\subsection{Effect of Rank 1} 
\label{ssub:interleave_sensitivity}

Owing to the 3 distinct classes of retrieval models, we find that each class selects a different document at rank 1. For language model based approaches, the document with highest textual relevance is at rank 1 while for the diversification based approaches marginal utility of the aspects decides the order of the ranking. One can argue that the effectiveness of the retrieval model may lie solely in how well it can rank the first document. Notice that no retrieval model that starts from a worse document is better than retrieval models from the other classes in terms of coverage. To address this issue and further establish the efficacy of combining diversity and interleaving, we devised an experiment where all retrieval models start with the same document - the document with the highest textual relevance to the user's initial query. 

In this experiment we consider the approaches in Figure \ref{fig:sfig_lm}.  Interleaving is employed specifically to prevent specialization of a query aspect. However as seen in Section \ref{sub:extrinsic_measures}, interleaving is sensitive to the choice of the initial ranking used. However if we use the language model for both the dynamic and static ranking we observe little to no improvement in coverage (see \interleavelmlm). 


Using either keyphrases or entities to diversify while interleaving improves coverage. Following from our previous discussion, entity based diversification outperforms keyphrase based diversification significantly. In fact \interleaveLmDivEntFeedback{} achieves the highest disambiguation accuracy across all approaches, the second highest coverage and also has high engagement. By interleaving two contrasting rankings we quickly obtain important keyphrases resulting in high disambiguation accuracy and also avoid long stretches of inconsequential documents resulting in high engagement. 

\textbf{Robustness.} Lastly we consider the robustness of the various approaches (Table \ref{tab:winloss}). \interleaveLmDivEntFeedback{} is the most robust approach for the given workload. For entities that have very few relevant documents in the collection, the textually relevant results might prove to be sufficient to gather enough keyphrases. Hence interleaving with \lm{} makes our approach more robust. \interleaveLmDivKpFeedback{} is also comprehensively more robust than its' closest rival \interleavedivKpKpFeedback.  


\subsection{User Study} 
\label{sub:user_study}

The entity addition task for long-tail ambiguous entities is a specialized task. Crowd-sourcing such a task can be unwieldy since pooling is infeasible, the any-time abandonment is hard to control and the number of approaches is large. To overcome this, we conducted a controlled lab experiment with 3 users, 3 distinct retrieval models from each class in Section~\ref{sub:baselines} (\,\lm{}, \interleaveLmDivKpFeedback{} and \interleaveLmDivEntFeedback{}\,) and 15 queries. The users are presented with a system similar to Fig.~\ref{fig:ui}~\cite{Hoffart:knowledgeAwakens}. They are also asked to read the Wikipedia page related to the entity being added before starting the task. Users are instructed to evaluate the first 20 documents they encounter for each entity. We asked the users to only judge if a document is relevant or not. We rely on the ground truth keyphrases from \texttt{EntityKB} to select the relevant keyphrases from a document for the user.

Table~\ref{tab:acc_user} shows the disambiguation accuracy results from the user study. We found a similar trend here as compared to the user simulation. \interleaveLmDivEntFeedback{} achieved the highest disambiguation accuracy -- close to 42\% at k$=20$. At the same depth \lm{} and \interleaveLmDivKpFeedback{} both achieved 35\% accuracy. In terms of coverage we found an interesting result: even though the coverage of \interleaveLmDivKpFeedback{} was consistently lower than \lm{} for $k>5$ ($Cov@20 = 0.432$ for \lm{} and $0.402$ for \interleaveLmDivKpFeedback{}) they both achieved very similar disambiguation accuracy. This implies \interleaveLmDivKpFeedback{} is able to cover just as many important discriminative keyphrases as \lm{} in spite of suffering from low overall coverage for real users. For engagement we once again found that \interleaveLmDivEntFeedback{} performs the best.

\begin{table}[t]
  \footnotesize
  \centering
   \scalebox{0.85}{
  \begin{tabular}{@{}llcccc@{}}\toprule

    \multicolumn{2}{l}{} & 5 & 10 & 15 & 20\\
    \midrule

    \multicolumn{2}{l}{\lm} &21.11\%&29.60\%&35.22\%&35.50\%\\
    \multicolumn{2}{l}{\interleaveLmDivKpFeedback} &20.01\%&31.43\%&31.98\%&35.45\%\\
    \multicolumn{2}{l}{\interleaveLmDivEntFeedback} & 24.03\%&30.00\%&39.47\%&41.44\%\\

    \bottomrule
  \end{tabular}}
  \caption{Disambiguation accuracy results for the user study at $k=5,10,15,20$.}
  \label{tab:acc_user}
\end{table}



\textbf{Summary: }
In this section we have empirically shown, with a user study and simulations, that entity based diversification approaches are better suited to the task of context gathering for ambiguous long tail entities. \interleavedivEntEntFeedback{} performs the best in terms of keyphrase coverage. We found that incorporating user feedback actively is beneficial especially for \lm{} in terms of coverage and engagement but only marginally in disambiguation accuracy. Adding user feedback to the diversification approaches causes slight improvement overall. Interleaving is useful in improving coverage for both keyphrase and entity based diversification approaches. When we interleave two contrasting approaches like in the case of \interleaveLmDivEntFeedback{} we also see a significant improvement in robustness and disambiguation accuracy. We also found that our approach is effective even when the document at rank 1 is the same for all competing approaches. Overall \interleaveLmDivEntFeedback{} achieves the best balance between coverage, engagement, disambiguation accuracy and robustness for the task of adding entities in context.

\note{
\begin{enumerate}
  \item ideal curve - how did we get it? max 80 percent - ideal docs cannot be ranked using text relevance
  \item why different starting position for each type of ranking (ranking with same starting position is important try the exp)
  \item 1. no feedback methods worse than feedback methods - lm feedback better than lm (give numbers - percent increase)
  \item kpdiv also improved with feedback 
  \item 2. effect of interleaving - interleaving helps? which combo helps? grad interleaving is better than static. feedback vs feedback+interleaving ---we need lm +lm-feedback with interleaving
  \item interleaving is sensitive to choice of retrieval model - choose diff models - lm +lm or div+div sucks
  \item Our approach - ent based div is good, growing always, stat significance , example where it breaks the flat coverage
\end{enumerate}
\paragraph{Engagement}
\begin{enumerate}
  \item low coverage - high dropout likelihood
  \item lm in the beginning may cause the user to dropout. feedback improves things.
\end{enumerate}
}



\section{Related Work}
\label{sec:related_work}

There is ample work on automatically identifying new or emerging entities. This task has been part of the TAC Knowledge Base Population track \cite{Ji:2011un} since its inception. Here, mentions referring to entities that are not part of the knowledge base should be identified and clustered by meaning. These clusters could in principle be added to a KB as new entities, but the precision of about 75\% \cite{TACKBP:2015} is still not nearly high enough to do this without human supervision. Other works have focused on extending existing entities with new keyphrases mined from a collection \cite{Li:2013vw,Hoffart:2014hp}. 


A natural application where users need entities going beyond Wikipedia-based knowledge bases is entity-based search. Here, the goal is to retrieve documents linked to KBs by querying for contained entities or categories \cite{Dalton:2014ct,Bast:2014vi,Hoffart:2014dt}. To the best of our knowledge, our work is the first to propose a retrieval-assisted \emph{manual entity addition} for high quality entity representations for emerging and long-tail entities.  

\textbf{Named Entity Disambiguation Systems:}
There is a recent survey on named entity disambiguation (also called entity linking) systems~\cite{Shen:2015wi}.  It contains a list of all methods that use the textual context of a mention as feature for the disambiguation process, the feature that our work is also focusing on. Together, these methods represent a large fraction of all methods discussed, which shows the wide applicability of the method. In principle, other features can be mined from user-accepted documents as well, however we do not explicitly discuss this extension in this paper.

\textbf{Search Result Diversification:}
To gather a well rounded representation of the entity we may present the user with diverse documents. Traditional diversification retrieval models are designed to satisfy ambiguous or multi-faceted queries. The key to diversification in either case lies in identifying query subtopics or aspects from the underlying information space accurately and then maximizing coverage of these aspects in the top-k results~\cite{agrawal2009diversifying}. For diversifying web results, various types of aspects have been considered: ~\cite{dou2011multi} uses anchor text mined from pseudo-relevant documents, query logs and website clusters; ~\cite{Santos:2010:EQR:1772690.1772780} uses query suggestions from a commercial search engine; ~\cite{agrawal2009diversifying} considers concepts from the Open Directory Project, to name a few. We choose to model the underlying information space as a set of keyphrases and entities mined from the documents since it is more natural to our task.

\textbf{User Feedback:}
Effectively utilizing the human in the loop requires careful consideration of the relevance feedback. Typically, relevance feedback can be gathered explicitly by asking the user to judge documents, implicitly using logs or automatically from pseudo relevant documents~\cite{harman1992relevance}. This feedback is then used to expand the query, most commonly using Rocchio's Algorithm~\cite{rocchio1971relevance} or \cite{lavrenko2001relevance}. The difficulty in query expansion lies in selecting appropriate terms to expand the original query. Previous work has looked at mining expansion terms from external sources like user logs~\cite{cui2003query} as well as more document and collection centric approaches like \cite{xu1996query,xu2000improving}. In a typical search engine explicit feedback is not used since gathering sufficient feedback for effective expansion is expensive~\cite{manning2008introduction}. However in our scenario, the user must provide explicit feedback on a document and keyphrase level on the fly. In our approaches we expand the query incrementally using only the relevant keyphrases judged by the user instead of mining new terms.

\section{Conclusion}
\label{sec:conclusion}

In this paper, we propose a retrieval-based approach for aiding addition of named 
entity representations using the human in the loop. We address the problem of 
reducing user effort, while ensuring high engagement, in identifying 
descriptions for ambiguous long-tail entities. We devise methods to incorporate 
the user feedback obtained during the addition process and identify the problems of specialization and concept drift. We propose different diversification approaches to maximize the coverage of relevant keyphrases and interleaving techniques to ensure engagement and robustness. We conducted extensive experiments, using real users and a simulation, showing that our approaches convincingly outperform carefully selected baselines in both intrinsic and extrinsic measures while keeping the users engaged.
Specifically, we find that diversifying over the entity space while taking feedback into account (\,\divEntFeedback{}) interleaved with \lm{} is the best performing approach in both intrinsic as well as extrinsic measures. In our future work we would like to take into account human errors in the entity addition process.

\section{Acknowledgment}

This work was carried out in the context of the ERC Grant (339233) ALEXANDRIA.


{ \small

}
%
%

\end{document}